# Microlensing Search of $10^6$ Quasars

**Andrew Gould**[1]

Dept of Astronomy, Ohio State University, Columbus, OH 43210

e-mail gould@payne.mps.ohio-state.edu

## Abstract

By monitoring $10^6$ quasars one could search for lensing by stars and Massive Compact Halo Objects (Machos) out to redshifts $z \sim 4$. If Machos have a present cosmological density $\Omega_{L,0} = 1\%$, then the expected event rate is $\Gamma \sim 200\,\mathrm{yr}^{-1}$. The expected event rate for known stars in galaxies is $\Gamma \sim 20\,\mathrm{yr}^{-1}$ assuming that their present cosmological density is $\Omega_{L,0} = 0.3\%$. Typical event times are $t_e \sim 3\,\mathrm{yr}$ for Machos and $t_e \sim 10\,\mathrm{yr}$ for stars. By comparing the optical depths to quasars at different redshifts, one could measure the star-formation and Macho-formation history of the universe. By comparing the time scales of events found parallel and perpendicular to the Sun's motion relative to the microwave background, one could measure or constrain the characteristic scale of large scale motions. The lensing events themselves would help probe the structure of quasars on scales of 50–1500 AU. The monitoring program could be carried out with a single dedicated 1 m telescope with a 4 deg$^2$ camera. Quasar lensing events can be unambiguously distinguished from quasar variability because in the former case the broad lines are unaffected while in the latter they respond to the variation in the continuum on times scales $\sim 1\,\mathrm{yr}$.

Subject Headings: gravitational lensing – large-scale structure of the universe – quasars: general – stars: masses

## 1. Introduction

The MACHO collaboration (Alcock et al. 1993, 1995b) and the EROS collaboration (Aubourg et al. 1993) have found a total 5 candidate lensing events toward the Large Magellanic Cloud (LMC). MACHO (Alcock et al. 1995a; Bennett et al. 1995) and the OGLE collaboration (Udalski et al. 1994) have found a total of 56 candidate events toward the Galactic bulge. From the time-scales of the bulge events, one can infer that the mass distribution is roughly consistent with what one would expect from stars in the solar neighborhood (Han & Gould 1995). On the other hand, the LMC events may be due to sub-stellar objects. First estimates indicate that the total mass of such Massive Compact Halo Objects (Machos) in the Galaxy may be of order $\sim 20\%$ of the dark halo (Alcock et al. 1995b), corresponding to a fraction of the closure density $\Omega \sim \mathcal{O}(1\%)$. In any event, these initial results demonstrate that microlensing is a powerful technique for finding stellar and sub-stellar objects solely from their gravitational effect on light.

To obtain a comprehensive picture of the distribution of stellar and near-stellar mass objects, one would like to apply the microlensing technique beyond the neighborhood of the Milky Way. The problem is that source stars at greater distances are generally not resolved. Crotts (1992) and Baillon et al. (1993) have shown that one can monitor unresolved stars in M31 by "pixel lensing" and have initiated observational programs to do so. I have shown that using the *Hubble Space Telescope* (*HST*) it is possible to detect Virgo intra-cluster Machos from pixel lensing of M87 (Gould 1995b). However, if one wishes to escape the narrow confines of the Local Supercluster, brighter sources than stars are required. Two possibilities are gamma ray bursters (GRB) (assuming that they are at cosmological distances) and quasars. Lensing of GRBs has been proposed as a method to detect a cosmological abundance of objects on many scales: $10^{-16} - 10^{-13}\,M_\odot$ from femtolensing (Gould 1992), $10^{-7} - 10^{-15}\,M_\odot$ from space-based parallaxes (Nemiroff & Gould 1995), $10^6 - 10^{11}\,M_\odot$ from time delays (Paczyński 1986, 1987; Blaes & Webster 1992; Mao & Paczyński 1992). No method for using GRBs to detect near-stellar sized objects has yet been proposed.

Dalcanton et al. (1994) have used quasars to probe for a cosmological abundance of lenses in the mass range $10^{-3} - 10^3 \, M_\odot$. The idea is that objects in this mass range would have Einstein rings large enough to cover the continuum region of a quasar, but not large enough to cover the broad-line region. Hence one would see a quasar with anomalously weak emission lines. More precisely, one should see more such weak line objects in a sample of distant quasars (which have a higher optical depth to microlensing) compared to a sample of nearby ones. The absence of such a signal allowed Dalcanton et al. (1994) to place upper limits on the abundance of these objects.

Here I present a practical program to monitor $N \sim 10^6$ quasars as a method to search for microlensing by near-stellar mass objects. I show that a fraction $\tau \sim 0.2 \, \Omega_{L,0}$ of these quasars should be lensed at any given time. Here $\Omega_{L,0}$ is the *present* density of lenses as a fraction of the critical density of the universe. This means that if $\Omega_{L,0} \sim 1\%$ in Machos, then a total $N\tau \sim 2000$ of these quasars would be lensed at any time. Just based on the known populations of stars in galaxies $\Omega_{L,0} \sim 0.3\%$, so $N\tau \sim 600$. For Machos, the typical event has a characteristic Einstein crossing time $t_e \sim 3 \, \mathrm{yr}$ while for stars, $t_e \sim 10 \, \mathrm{yr}$. The total event rate would be $\Gamma \sim 200 \, \mathrm{yr}^{-1}$ for Machos of mass $M = 0.5 \, M_\odot$ and $\Gamma \sim 20 \, \mathrm{yr}^{-1}$ for stars. The optical depths and event rates are about double for an open ($\Omega = 0$) compared to a closed ($\Omega = 1$) universe. The values quoted above are an average of the two.

A search for microlensing of quasars would provide a powerful probe of cosmological models, of the star-formation and Macho-formation history of the universe, and of quasars themselves. Here I briefly outline how such a program might be carried out, what might be learned, and what systematic effects and backgrounds it would have to overcome.

## 2. The Search

The surface density of quasars to $B = 22$ is $100\,\mathrm{deg}^{-2}$ (Hartwick & Schade 1990). For typical quasars, $B - V \sim 0$. The dark sky has a surface brightness of $V = 22\,\mathrm{mag\,arcsec}^{-2}$. For isolated sources and assuming $1''$ seeing, a 1 m telescope (25 photons s$^{-1}$ at $V = 20$) could therefore achieve signal to noise of 15 in a 4 minute exposure. Allowing 1 minute for pointing and readout, 110 such exposures could be made in an average 9 hour night, or 1650 in 15 equivalent dark nights per month. Assuming 25% bad weather (conditions need not be photometric) this would imply 1250 exposures per month. There is no serious problem constructing a 1 m telescope with a 2 deg field of view. MACHO already has two operating cameras covering $0.5\,\mathrm{deg}^2$ with $0.''6$ pixels. EROS is close to completing a camera that is twice as big. Thus, a $4\,\mathrm{deg}^2$ camera is easily within reach. Hence, one could image $5000\,\mathrm{deg}^2$ per month to $V = 22$ at S/N=15 with only a modest investment in equipment. If every field were monitored for 6 months per year, a total of $10^4\,\mathrm{deg}^2$ would be covered. These would contain a total of $10^6$ quasars. There are only $2 \times 10^4\,\mathrm{deg}^2$ of high latitude ($|b| > 30°$) fields in the entire sky, so this would be a significant fraction of all available fields. For a telescope located relatively near the equator (e.g. Hawaii) there would be two 4-hour gaps in the object list near the Galactic plane, implying that at times the observations would have to be made 2 hours over. This would not be a severe restriction. It is possible that some fields would be missed during some months due to a combination of bad weather and bad field position (especially the Fall fields) but since the event times are of order several years, this problem is not critical.

The detection of the microlensing events could be carried out in one of two ways. Probably a combination of both should be used. First one could form a catalog of every object $V < 22.5$ and measure its flux in each observation. Gould, Bahcall, & Maoz (1993) found 186 stars $V < 21.3$ in an ensemble of 143 fields ($|b| > 30°$) covering $0.047\,\mathrm{deg}^2$. Using this result and extrapolating it to $V = 22.5$ (with a $\gamma = 0.35$ luminosity function power law) I estimate that the combined fields contain a total of $\sim 6 \times 10^7$ stars. This is comparable to the total number of stars being monitored by MACHO, but with many fewer observations per year (6 versus $\sim 100$). Hence the data-base problems are within the scope of what is already being done. No attempt would be made to distinguish quasars from other objects. The subset of objects that varied would be analyzed to see which might be quasar lensing candidates. In this way the search would be very similar to the Galactic microlensing searches now being carried out by MACHO, EROS, and OGLE.

The point-source approach described above would work very well for isolated quasars, which comprise the vast majority of quasars at high latitude. However, most of the known stars in the universe are in galaxies, and most of the quasars that get lensed by these known stars will therefore have projected positions close to galaxies. It would require a gigantic effort to distinguish all of these quasars from the more numerous globular clusters, HII regions, and other structures. Fortunately, this is also unnecessary. Instead, one would use the techniques now being developed for pixel-lensing searches to find point-source variations behind galaxies. Two methods are now being tested and refined. Crotts (1992) convolves his best image with the seeing of the current image and subtracts this convolution from the current image. He then searches the difference image for point sources. Baillon et al. (1993) look for flux changes in individual pixels. A third method would be to construct a median image out of a series of images with similar seeing and then subtract that from the individual images. In any event, it should not be difficult to apply the pixel-lensing methods to finding varying quasars behind galaxies.

## 3. Optical Depth and Event Rate

To estimate the optical depth and event rate, I assume that the density of lenses is given by

$$\Omega_L(z) = (1+z)^{-\alpha}\Omega_{L,0}, \tag{3.1}$$

where $\Omega_{L,0}$ is the density of lenses today as a fraction of the critical density, and $\alpha$ parameterizes the history of lens (i.e., star or Macho) formation. For example, $\alpha = 0$ corresponds to all lenses forming before the epoch of quasars. If the lens formation rate is uniform in time, then $\alpha = 3/2$ for an $\Omega = 1$ universe. For simplicity of exposition, I consider only two geometries: $\Omega = 0$ and $\Omega = 1$ with a "filled beam" ($\beta = 1/4$ in the notation of Schneider, Ehlers & Falco 1992). In the latter model, most of the matter in the universe is assumed to be smooth and lenses constitute a small perturbation of lumpiness. (Note however, that the results for an $\Omega = 1$ "empty beam" universe, which is appropriate for $\Omega_{L,0} \sim 1$ differ by only $\sim 20\%$ from the "filled beam".) For these two geometries, the angular diameter distance between redshifts $z_1$ and $z_2$ are (Schneider et al. 1992)

$$D(z_1, z_2) = \frac{1}{2}\frac{c}{H_0} w_1(w_1^{-2} - w_2^{-2}), \qquad (\Omega = 0), \tag{3.2}$$

and

$$D(z_1, z_2) = 2\frac{c}{H_0} w_2^{-1}(w_1^{-1/2} - w_2^{-1/2}), \qquad (\Omega = 1), \tag{3.3}$$

where $w_{1,2} \equiv 1 + z_{1,2}$. As usual, the Einstein radius $r_e$ is given by

$$r_e^2 = \frac{4GM D_{\text{OL}} D_{\text{LS}}}{c^2 D_{\text{OS}}}, \tag{3.4}$$

where $M$ is the mass of the lens and $D_{\text{OL}}$, $D_{\text{LS}}$, and $D_{\text{OS}}$ are the angular diameter distances between the observer, lens, and source. The density of lenses as a function of redshift is given by $n(z)/n(0) = (1+z)^{3-\alpha}$ taking account of both the expansion of the universe and the formation of lenses. Hence, the differential optical depths

(see Turner, Ostriker, & Gott 1984, but note the correction for the "filled beam") are

$$\frac{d\tau}{dx} = \frac{3}{4}\Omega_{L,0}\frac{(y^2 - x^2)(x^2 - 1)}{(y^2 - 1)}x^{-2-\alpha}, \qquad (\Omega = 0), \tag{3.5}$$

and

$$\frac{d\tau}{dx} = 3\Omega_{L,0}\frac{(y^{1/2} - x^{1/2})(x^{1/2} - 1)}{(y^{1/2} - 1)}x^{-3/2-\alpha}, \qquad (\Omega = 1), \tag{3.6}$$

where $x = 1 + z_L$ and $y = 1 + z_Q$. These yield total optical depths

$$\tau = \frac{3}{2}\frac{1}{(1-\alpha)(y^2 - 1)}\left(\frac{y^{3-\alpha} - 1}{3 - \alpha} - \frac{y^2 - y^{1-\alpha}}{1 + \alpha}\right), \tag{3.7}$$

and

$$\tau = \frac{3}{2}\frac{1}{\alpha(y^{1/2} - 1)}\left(\frac{y^{1/2} - y^{-\alpha}}{\alpha + 1/2} - \frac{1 - y^{-\alpha+1/2}}{\alpha - 1/2}\right). \tag{3.8}$$

Figure 1 shows the optical depths as functions of quasar redshift for an open and closed universe with various values of $0 \leq \alpha \leq 2.5$. For present purposes, the important point is that for typical quasar redshifts $z_Q \sim 2$ and a plausible estimate $\alpha \sim 3/2$, the optical depth is $\tau/\Omega_{L,0} \sim 0.15$ for $\Omega = 1$ and twice as great for $\Omega = 0$. Adopting $\Omega_{L,0} = 0.3\%$ for stars in galaxies, I find $\tau \sim 5 - 10 \times 10^{-4}$, so that if $N \sim 10^6$ quasars are monitored, then $\sim 500 - 1000$ of them will be lensed by stars at any given time. If there is an additional $\Omega_{L,0} = 1\%$ in Machos in the halos of galaxies, then there are an additional $\sim 1500 - 3000$ quasars lensed by these.

To calculate the event rate $\Gamma$, I first estimate the typical mass. For stars I base my estimate on the mass function of nearby stars which peaks at $M \sim 0.25\,M_\odot$ when expressed as number per unit mass and at $M \sim 0.5\,M_\odot$ when expressed per log mass (Gould, Bahcall, & Flynn 1995). For gravitational lensing (where $\Gamma \propto r_e \propto M^{1/2}$), an appropriate choice is therefore $M \sim 0.35 M_\odot$. However, as will be seen below, typically the Einstein ring is $\sim 1000\,\mathrm{AU}$. Many stars have binary companions well within this radius. I therefore adopt $M \sim 0.5\,M_\odot$ for stars. For Machos I adopt $M \sim 0.05\,M_\odot$ which is consistent with the MACHO and EROS data.

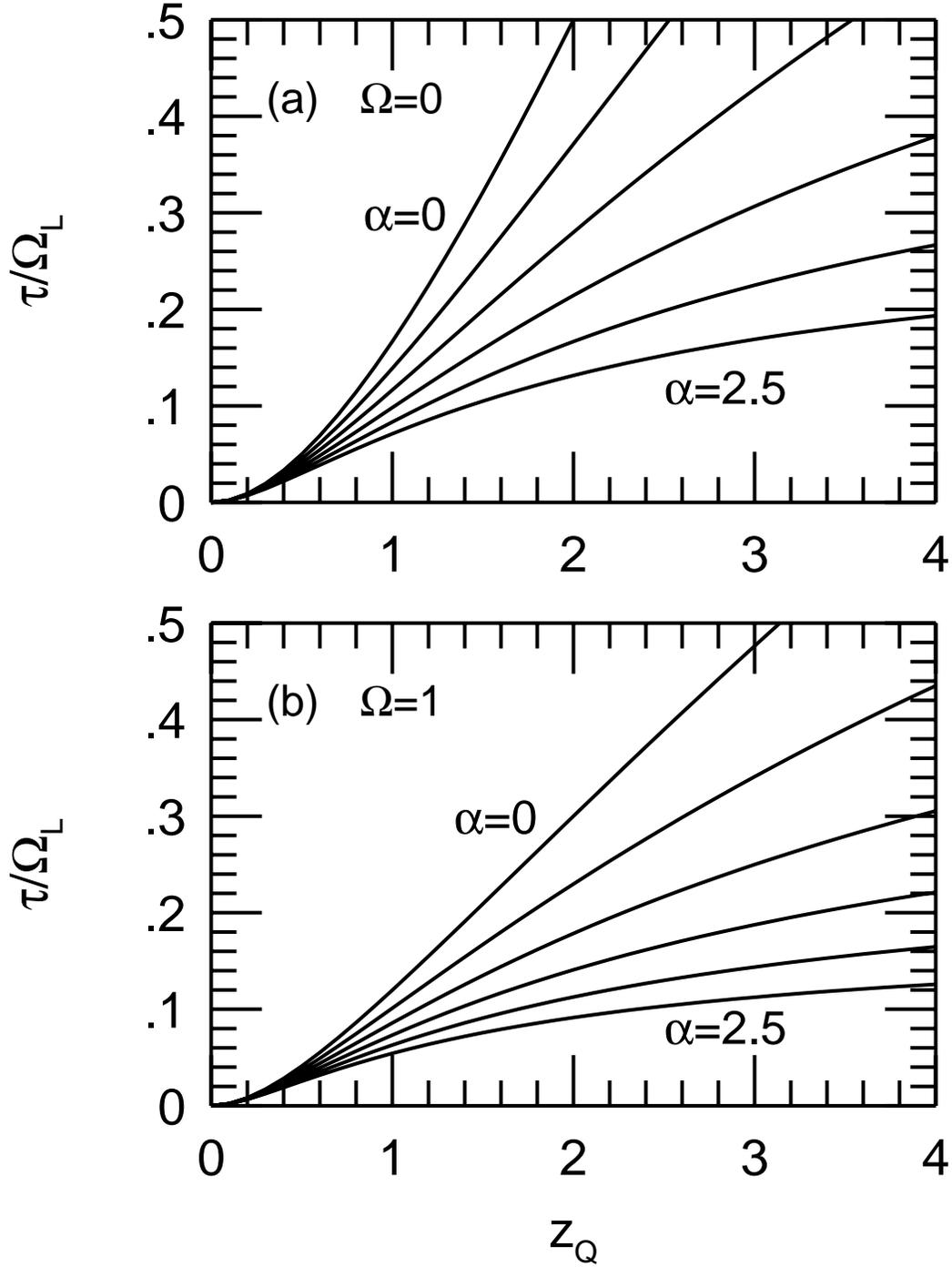

Figure 1. Optical depth ($\tau$) per unit present-day density of lenses ($\Omega_L$) as a function of quasar redshift ($z_Q$) for $\Omega = 0$ (a) and $\Omega = 1$ (b) universes. Curves are for star-formation parameters $\alpha = 0, 0.5, 1, 1.5, 2$, and $2.5$. The fraction of stars scales with redshift $z_L$ as $(1+z_L)^{-\alpha}$.

For simplicity, I assume that all of the quasars are at redshift $z_Q = 2$. With $M$ and $z_Q$ fixed, the Einstein radius depends only on $z_L$ (for a given geometry). To find $\Gamma$, it is then only necessary to calculate the distribution of transverse velocities at each redshift. The crossing time is then given by

$$t_e = \frac{r_e}{v}, \qquad \mathbf{v} = x^{-1}\mathbf{v}_L - \frac{D_{\text{LS}}}{D_{\text{OS}}}\mathbf{v}_\odot - y^{-1}\frac{D_{\text{OL}}}{D_{\text{OS}}}\mathbf{v}_Q, \qquad (3.9)$$

where $\mathbf{v}_L$, $\mathbf{v}_Q$, and $\mathbf{v}_\odot$ are respectively the transverse (2-D) velocities of the lens, quasar, and Sun relative to the microwave background (CMB). The last of these is known very precisely, $v_\odot = 370 \pm 3\,\text{km s}^{-1}$ toward $l = 264.4 \pm 0.2$, $b = +48.1 \pm 0.4$ (Bennett et al. 1994). The velocity distributions of the lens and quasar are of course not known. I estimate them as follows. I assume that the present-day velocity distribution of galaxies relative to the CMB is primarily the result of linear perturbations and is Gaussian distributed with 1-dimensional dispersion $\Delta$. Two values of $\Delta$ will be considered: $\Delta = 350\,\text{km s}^{-1}$ (which assumes that the Local Group speed is typical: $3^{1/2}\Delta \sim 600\,\text{km s}^{-1}$), and $\Delta = 175\,\text{km s}^{-1}$. Linear perturbations evolve with time, shrinking by $g_\Omega(x)$ at higher redshift, with $g_1(x) = x^{-1/2}$ and $g_0(x) = x^{0.6}$. I also assume that the velocities of stars in galaxies result from non-linear perturbations and have a gaussian distribution with dispersion $\zeta = 100\,\text{km s}^{-1}$. These do not evolve in time. I assume that the quasars are sitting at the bottom of deep potential wells and therefore are not affected by these non-linear motions. Combining these assumptions, I find that the transverse velocity distribution is given by

$$f(v, \theta) = \frac{1}{2\pi\sigma^2}\exp\left(-\frac{v^2 + v_0^2 - 2vv_0\cos\theta}{2\sigma^2}\right), \qquad (3.10)$$

where

$$\sigma^2 = \left\{\left[\frac{g_\Omega(x)}{x}\right]^2 + \left[\frac{g_\Omega(y)}{y}\frac{D_{\text{OL}}}{D_{\text{OS}}}\right]^2\right\}\Delta^2 + \frac{\zeta^2}{x^2}; \qquad v_0 = \frac{D_{\text{LS}}}{D_{\text{OS}}}v_\odot. \qquad (3.11)$$

Hence, the transverse speed distribution is given by

$$f(v) = \frac{v}{\sigma^2} \exp\left(-\frac{v^2 + v_0^2}{2\sigma^2}\right) I_0\left(\frac{v v_0}{\sigma^2}\right), \qquad (3.12)$$

where $I_0$ is a modified Bessel function. The event rate can then be expressed as a function of event time, $t_e \equiv r_e/v$, as

$$\frac{d\Gamma}{dt_e} = \frac{2}{\pi} N \int_1^y dx \frac{d\tau}{dx} r_e^{-1}(x) \int dv\, v f(v) \delta\left[t_e - \frac{r_e(x)}{v}\right] = \frac{2}{\pi} \frac{N}{t_e^2} \int_1^y dx \frac{d\tau}{dx} F\left[\frac{r_e(x)}{t_e}\right], \qquad (3.13)$$

where $F(v) \equiv v f(v)$ is dimensionless.

Figure 2 shows the event rates for eight different models. Figures 2(a-b) show rates for models with $\Omega_{L,0} = 1\%$ in Machos with $M = 0.05\,M_\odot$, and with $\Delta = 350\,\mathrm{km\,s^{-1}}$ and $\Delta = 175\,\mathrm{km\,s^{-1}}$ respectively. The solid lines assume that all fields are perpendicular to the Sun's direction of motion relative to the CMB ($v_\odot = 350\,\mathrm{km\,s^{-1}}$) while the dashed lines assume that all fields are parallel to this motion ($v_\odot = 0$). Both $\Omega = 0$ and $\Omega = 1$ universes are shown. Figures 2(c-d) are similar except that $\Omega_{L,0} = 0.3\%$ in stars with $M = 0.5\,M_\odot$. In all cases, $\alpha = 1.5$, $H_0 = 75\,\mathrm{km\,s^{-1}\,Mpc^{-1}}$ ($h = 0.75$), and $z_Q = 2$. I find that most of the optical depth occurs for lens redshift $0.25 \lesssim z_L \lesssim 1$ for which $r_e \sim 1300\,\mathrm{AU}$, confirming that binary companions should normally be included when estimating the mass of the lens.

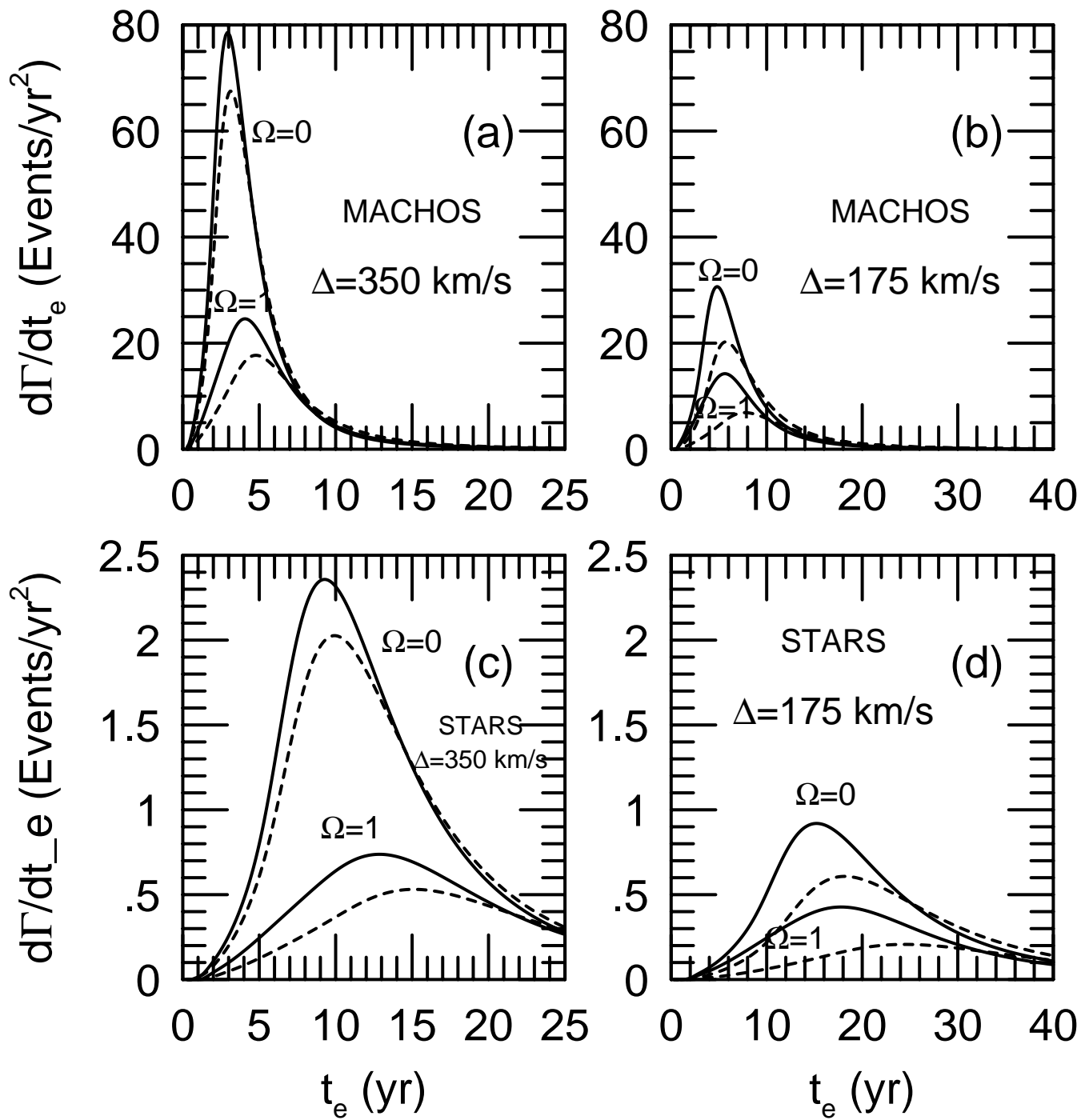

Figure 2. Event rates ($\Gamma$) per unit crossing time ($t_e$) per year of observation, assuming all fields are perpendicular (*solid*) or parallel (*dashed*) to the Sun's CMB motion. Panels (a) and (b) assume a cosmological density of Machos $\Omega_{L,0} = 1\%$ with mass $M = 0.05\,M_\odot$. Panels (c) and (d) assume $\Omega_{L,0} = 0.3\%$ in stars with mass $M = 0.5\,M_\odot$. Panels (a) and (c) assume a 1-dimensional present-day cosmic dispersion $\Delta = 350\,\mathrm{km\,s^{-1}}$. Panels (b) and (d) assume $\Delta = 175\,\mathrm{km\,s^{-1}}$.

## 4. Applications

By monitoring $10^6$ quasars, one would be able to address a number of important questions about cosmology, about stars and Machos, and about quasars.

Most directly, one would find the cosmological abundance and approximate mass range of stellar and near-stellar mass objects. For example, for an open universe, $\Omega_{L,0} = 1\%$ in $M = 0.05\,M_\odot$ Machos, and $\Delta = 350\,\text{km}\,\text{s}^{-1}$, the total event rate would be $\Gamma \sim 290\,\text{yr}^{-1}$ with a fairly sharp peak near $t_e \sim 3\,\text{yr}$ (see Fig. 2a). For a closed universe, the event rate is substantially lower, $\Gamma \sim 140\,\text{yr}^{-1}$, implying that the estimate of the density of the lenses is somewhat dependent on $\Omega$. However, the optical depth ($\tau = 0.22\%$ vs. $\tau = 0.14\%$) is less sensitive to $\Omega$. Both the closed and open universes have peaks near $t_e \sim 3\,\text{yr}$, indicating that within the framework of a fixed $\Delta$, the characteristic mass of the lenses can be determined from the characteristic times. Note that this signal is substantially larger than that currently being obtained by MACHO and EROS in microlensing searches toward the LMC. Moreover, it is possible that there is a significant population of Machos not associated with galaxies (e.g. formed in the Lyman $\alpha$ clouds) in which case the signal would be even larger. However, it is also possible that there are no Machos and that stars in galaxies are the only substantial population of near stellar-mass objects. In this case (Fig. 2c), the signal would be much weaker, $\Gamma \sim 28\,\text{yr}^{-1}$ ($\Omega = 0$) and $\Gamma \sim 13\,\text{yr}^{-1}$ ($\Omega = 1$), with optical depths $\tau \sim 0.06\%$ and $\tau \sim 0.04\%$. This is still not negligible. One might be concerned that the typical stellar events $t_e \sim 10\,\text{yr}$ would not be recognizable as such until many years into the experiment. However, it should be noted that a fraction $\beta$ of all events have impact parameters $b < \beta r_e$ and that these events have "effective peak times" $t_{peak} = (b/r_e)t_e < \beta r_e$. Hence, even during the first year there would be $\sim 2$ recognizable events and the total would grow quadratically for several years.

Second, one could obtain information about the star-formation (or Macho formation) history of the universe from the distribution of optical depths as a function of quasar redshift $\tau(z_Q)$. Notice from Figure 1(a) ($\Omega = 0$) that for $\alpha = 0$ $\tau(3)/\tau(1) = 5.4$ while for $\alpha = 1.5$, $\tau(3)/\tau(1) = 3.1$. For $\Omega = 1$ (Fig. 1(b)), the corresponding ratios are 4.0 and 2.3. Hence, if for example Machos formed before a redshift of 3, then the growth of optical depth with quasar redshift would be much steeper than if they formed continuously. In the latter case, there would be few Machos at high redshift and hence high-redshift quasars would have little advantage over those at low redshift. Note that to interpret this slope as a measurement of $\alpha$, it is necessary to independently constrain $\Omega$.

Third, one would be able to obtain information about the transverse dispersion of galaxies, $\Delta$. Suppose that $\Delta = 175\,\mathrm{km\,s^{-1}}$ and $\Omega = 1$. Then from Figure 2(b), one sees that the number of short events ($t_e \lesssim 8\,\mathrm{yr}$) seen in the direction of the Sun's motion is less than ($\sim 1/3$) in the direction perpendicular to its motion. The ratio is larger ($\sim 2/3$) for $\Omega = 0$, but the effect is still pronounced. On the other hand, for $\Delta = 350\,\mathrm{km\,s^{-1}}$ the ratios are much closer to unity. In particular, for $\Omega = 0$ the two distributions are practically indistinguishable. For low $\Delta$ much of the transverse velocity is provided by the Sun's motion, so the directional effect is large. Note again that to fully interpret an observed directional effect, one would have to independently constrain $\Omega$. This is because for a high $\Omega$ universe the motions at $z \sim 0.5$ (where typical lenses are) are much slower than in a low $\Omega$ universe.

All of the above effects apply to stars as well as Machos. However, they are more difficult to measure because of poorer statistics.

Also, by studying the lines of sight to lensed quasars and comparing them to lines of sight to unlensed quasars, one could learn about the association of lenses with galaxies and with damped Lyman $\alpha$ and Lyman limit systems. It might be, for example that the longer events (stars) are strongly associated with visible galaxies, while the shorter events (Machos) are less associated.

The survey could also provide important information about the quasars themselves. The typical size of an Einstein ring is $\sim 1300\,\mathrm{AU}$ for stars and $\sim 400\,\mathrm{AU}$ for Machos. Projected onto the quasar, these sizes grow by $D_{\mathrm{OS}}/D_{\mathrm{OL}}$, typically a factor $\lesssim 1.5$. For definiteness, I consider projected Einstein radii of 2000 and 600 AU for stars and Machos respectively.

These sizes should be compared with the limits placed on the source size of 2237+0305 (Huchra's Lens) from microlensing caustics. Wambsganss, Paczyński, & Schneider (1990) derived a limit of $r < 200\,\mathrm{AU}$ by modeling the caustic network of the lens using a Salpeter mass function with a cut off at $0.1\,M_\odot$ and assuming a transverse speed of $600\,\mathrm{km\,s^{-1}}$. Rauch & Blandford (1991) confirmed this limit and argued that it is a factor $\sim 10$ too small to be consistent with their favored accretion disk model and a factor $\sim 3$ too small for a black body model. A Salpeter mass function places most of the mass near the lower limit. As noted above however, the mass function derived by Gould et al. (1995) from *HST* star counts peaks at $\sim 0.35\,M_\odot$. Recall that I adjusted this value upward to account for binaries within 1 Einstein radius. The same adjustment should apply to Huchra's Lens. Since $r_e \propto M^{1/2}$, one may guess that the size limit is increased by a factor $\sim 2$. Huchra's Lens lies in a direction almost perfectly parallel to the CMB dipole so that essentially all of the transverse motion comes from the lensing galaxy. This makes the estimate of $600\,\mathrm{km\,s^{-1}}$ already somewhat high for $\Delta = 350\,\mathrm{km\,s^{-1}}$, but for deriving a limit one should use the highest plausible value. I adopt $900\,\mathrm{km\,s^{-1}}$ and so raise the limit to $\sim 600\,\mathrm{AU}$. Huchra's Lens is estimated to have an unmagnified, dereddened apparent magnitude of $I \sim 18.5$ (Rauch & Blandford 1991, and references therein), and a redshift $z_Q = 1.645$. This is $\sim 2$ mag more intrinsically luminous than a "typical" $z_Q = 2$ quasar with $B = 21.5$ (assuming $B - I \sim 0.7$). Assuming that luminosity scales as radius squared, this leads to a limit $r \lesssim 250\,AU$ for a "typical" quasar in the sample. From this line of argument, I conclude that the projected Einstein ring will generally be larger than the quasar source, so that significant magnification will occur provided that the center of the source comes within the Einstein ring.

However, the fact that the source size and the Einstein ring are roughly comparable means that the lensing event will often probe the structure of the quasar. For $\sim 10\%$ of the stellar events and $\sim 25\%$ of the Macho events, the lens will come with $\sim 250\,\mathrm{AU}$ of the center of the quasar. If this is indeed the physical size of the source, the light curve will then deviate from the standard microlensing form. For these events, the quasar could be monitored at many wavelengths. It is possible that at higher energies the structure could be probed at even smaller distance scales.

## 5. Backgrounds

There are many objects in the sky other than microlensed quasars which vary with time. There are variable stars, novae, supernovae, and of course unmicrolensed quasars. Most of these objects can easily be distinguished from quasar microlensing events. The overwhelming majority of variable stars at high latitude, for example, vary on much shorter time scales than the several years typical of lensing events. Novae and supernova will often show up in or near galaxies (the expected location of microlensed quasars). However, these events peak on time scales of weeks, not years. Although these events will present something of a nuisance, they will not compromise the search. Indeed, one of the main collateral benefits of the search will be an enormous catalog of variable stars and supernovae.

Variation in quasars will present two types of problems. First, most if not all quasars vary on many time scales, including the several year time scales typical of microlensing. As I describe below, it is possible in principle to distinguish between intrinsic variation and microlensing. However the tests for microlensing are difficult and would be impractical if they had to be carried out on a very large number of candidate events. For the search to be practical, a class of relatively quiescent quasars would have to be identified. Or more precisely, a class non-quiescent quasars would have to be identified and excluded from the sample. Identifying such quasars would not be difficult. They would immediately come to attention as part of the search for time variable objects. The problem would be setting the threshold of variability high enough so that it would exclude most problem objects, but not so high that it would destroy the sample. Based on current knowledge of quasar variability one cannot know whether it is possible to establish such selection criteria. It will, however, be possible when EROS begins its supernova search which will be carried out very much along the lines of the lensing search described here, except over a much smaller area $[\mathcal{O}(10^2)\mathrm{deg}^2]$ and with deeper and more frequent exposures (M. Spiro 1995, private communication). It will therefore detect $\mathcal{O}(10^4)$ quasars if all quasars vary and it will provide very good data on their variability. If some quasars do not vary, or vary below the threshold of sensitivity, this will be apparent from a comparison of the quasars found by EROS with those identified by other means in the same area of the sky.

Assuming that the sample can be weeded of most spurious candidates, how can the true lensing events be separated from the remaining false signals? For a lensing event, the continuum region will be magnified but the broad line region will not. An outburst in the continuum region might initially mimic this behavior. However, over several years the light from the outburst would reach the broad line region and excite additional activity there as well. Thus by monitoring the spectrum during an event one could distinguish the outbursts from the lenses. Because the broad lines are $\sim 10,000\,\mathrm{km\,s^{-1}}$ wide, low resolution spectra would be adequate.

A second potential problem caused by variability is that the interpretation of real events could be compromised by intrinsic variation. This problem could also be minimized by monitoring the spectra.

## 6. Systematic Effects

There are two major systematic effects which could affect the completeness of the survey: regions of high lensing optical depth and dust.

The analysis of § 3 implicitly assumed $\tau \ll 1$ along all lines of sight. However, known stars are concentrated in galaxies and in the centers of galaxies, at least, this limit does not apply. Regions of high optical depth affect the analysis in two ways. First, the lenses within a region of $\tau > 1$ do not microlens at all. Instead they macrolens. Second, the lenses in regions of moderate optical depth $0.2 \lesssim \tau \lesssim 1$ do not generate events with isolated-lens characteristics. Instead they form caustic networks. For an isothermal sphere composed of point masses, with rotation speed $v_c$, the optical depth is $\tau = \pi(v_c/c)^2(D/a)$ where $D = D_{\rm OL}D_{\rm LS}/D_{\rm OS}$ and $a$ is the impact parameter. For a typical case with the quasar at $z_Q = 2$ and an intervening galaxy at $z = 0.5$, $D = 0.156(c/H_0)$ for $\Omega = 0$ and $D = 0.138(c/H_0)$ for $\Omega = 1$. Adopting the mean and assuming $h = 0.75$, I find

$$\tau \sim \left(\frac{v_c}{220\,{\rm km\,s^{-1}}}\right)^2 \left(\frac{a}{\rm kpc}\right)^{-1}. \tag{6.1}$$

This shows that the region corresponding to the bulge of the Milky Way will not microlens at all, while the region corresponding to the inner disk scale length or so will be affected by caustic networks. (If the total density of stars is $\sim 0.003$, as has been assumed in this paper, then much of the gravitating matter in the inner portions of galaxies is in a smooth component rather than microlenses. In these conditions, regions of moderate optical depth generate less severe caustic networks.) Hence, the number of isolated lens events due to known stars has been overestimated by a factor $\sim 2$.

On the other hand, the events generated by these regions of high and moderate $\tau$ will nonetheless be extremely interesting. First, events taking place within a few arcsec of a galaxy can reasonably be associated with the galaxy. Thus, the redshift of the lens can be measured. Second, it is possible to tell from the light curve whether the lens is approximately isolated or a caustic. This will give direct information about the density of lumpy material compared to smooth dark matter. Third, in cases where the quasar is so close to the center of the galaxy that it is macrolensed, there will still be observable events. The quasar itself will be noticed either because of its intrinsic variation or because of caustic microlensing of the separate images. In either case, it will be noticed that there are multiple images. Hence, a substantial sample of interesting macrolenses will be assembled.

Lensing events discovered in spiral galaxies can be used to measure the transverse velocities of the host galaxies, provided they are viewed simultaneously from a satellite in solar orbit at $\sim 30\,\mathrm{AU}$ (Gould 1995a).

Note that Machos are very little affected by problems of high optical depth. If Machos follow the $r^{-2}$ distribution of the mass, then most of them lie beyond $10\,\mathrm{kpc}$ from the center of the galaxy where $\tau \lesssim 0.1$.

Since dust is concentrated in galaxies, it will tend to obscure some of the quasars lying behind the galaxies. This will complicate the interpretation of the lensing statistics. However, a large uniformly selected sample of quasars including quasars behind galaxies, is probably the best way to study the effects of dust.

**Acknowledgements**: I would like to thank K. Griest, R. Pogge, D. Weinberg, E. L. Wright for stimulating discussions. This work was supported in part by grant AST 94-20746 from the NSF.